# Three-Dimensional Cloaking Device Operates at Terahertz Frequencies

Fan Zhou[1]*, Yongjun Bao[1]*, Wei Cao[2]*, Jianqiang Gu[2], Weili Zhang[2], and Cheng Sun[1†]

[1] Mechanical Engineering Department, Northwestern University, Evanston, IL 60208, USA
[2] School of Electrical and Computer Engineering, Oklahoma State University, Stillwater, OK 74078, USA

* These authors contributed equally to this work

† To whom correspondence should be addressed. E-mail: c-sun@northwestern.edu

The invisibility cloak has been a long-standing dream for many researchers over the decades. Recently, this subject has attracted considerable interest with the introduction of transformational optics, which provides a general method to design material distributions to hide the subject from detection (*1, 2*). By transforming space and light propagation, a three-dimensional (3D) object is perceived as having reduced number of dimensions, in the form of points, lines, and thin sheets, making it "undetectable" judging from scattered field (*2-4*). Since the first experimental demonstration using resonant metamaterials (*5*), a variety of cloaking devices have been reported at microwave and optical frequencies (*6-8*) while the Terahertz (THz) domain remains unexplored. Moreover, it should be noted that all the previous experimental demonstrations are performed in a two-dimensional (2D) waveguide configuration. Although those works represent a critical step in validating the concept of the invisibility cloak, one would expect the cloaking device to be realized in 3D with the ability to cloak an object of realistic size. This requires the construction of an optically large cloaking device with features much smaller than the wavelength. Fabricating 3D structures with aspect ratio close to 100:1 is obviously a challenging task.

Here, we report an experimental demonstration of a 3D THz ground plane cloak. The 3D cloaking device was fabricated using the projection microstereolithography (PμSL) technique (*9*). A high resolution optical image projected on the top surface of photocurable polymer resin (1,6 Hexanediol Diacrylate, HDDA) defines the shape of a solidified thin polymer layer. Using a dynamic mask (1400 x 1050 pixels), the projected optical image can be controlled electronically. Rapid fabrication of the 3D structure according to the designed geometry is accomplished in a layer-by-layer fashion. The cloaking device shown in Fig. 1(a) is fabricated using 200 layers with each layer 20 μm thick, for a total thickness of 4 mm. The refractive index distribution of the triangle-shaped cloaking device is obtained by quasi-conformal mapping for transverse magnetic waves (*4*). Sub-wavelength holes of varying dimension (***d***) in a square unit cell (***a***=85.2 μm) are used to obtain the desired permittivity profile under the effective media approximation. The grayscale of individual pixel can be adjusted so the holes can be fabricated with sub-pixel precision. The characteristic of such cloak design consisting of microscopic hole structures is evaluated using numerical simulation for a bumped surface without and with a cloak, as shown in Fig. 1(b) and (c), respectively. The simulation shows the magnetic field component in the z direction at 0.6 THz. The THz wave reflected from the uncloaked bump exhibits two distinct peaks and one minor peak in between. In contrast, the reflection from the cloaked bump shows the expected flat wavefront, as if the bump does not exist. Thus, the object placed underneath the reflective bump can be cloaked.

Reflection terahertz time-domain spectroscopy (THz-TDS) was employed to characterize the cloaking samples (*10, 11*). The photoconductive switch-based THz-TDS system was optically gated by 30 fs, 800 nm optical pulses generated from a self-mode-locked Ti:sapphire laser. The terahertz radiation emitted from a GaAs transmitter was spatially gathered by a hyperhemispherical silicon lens and then collimated into a parallel beam before entering the cloaking sample. The reflected terahertz signal through the sample was then detected by a mobile silicon-on-sapphire (SOS) receiver optically excited with fiber-coupled femtosecond pulses (*12*). A 1 mm wide aluminum slit was attached to the SOS detector in order to improve the spatial resolution of the measurements. The detector was scanned 10 mm away from the output interface to measure the spatial distribution of the terahertz wavefront (Fig. 1a). The frequency-dependent terahertz amplitudes at each spatial position are retrieved via the Fourier transform of the measured time-domain signals. The measured spectra map of the uncloaked bump and cloaked bump samples are plotted in Fig. 1(d) and (e), respectively. The horizontal and vertical axes represent scan positions and frequencies, while the color represents the amplitude of the spectra. As shown in Fig. 1(d), two distinct reflection peaks can be clearly observed across a broad frequency range, which is caused by the reflection at the surface of the bump. The measured peak positions are consistent with the numerical simulation peak positions, represented by superimposed white crosses. By contrast, in the spectral map of the cloak sample, the wavefront is relatively smooth with a single peak observed near the 0 mm position [Fig. 1(e)]. The discontinuity of the reflection peak near 0.32 THz is likely due to the multiple scattering within the triangle-shaped cloaking device.

**Figure caption:**

**Fig. 1. Three-dimensional invisibility cloak at terahertz frequency.** (a) Schematics showing design, fabrication, and characterization of 3D THz cloak. The refractive index profile of the triangle-shaped cloak structure is obtained using quasi-conformal mapping. The 3D cloak structure is fabricated using the projection microstereolithography technique. The surface of the cloaking device was metallized to enhance the contrast for better representation of the fine features in the scanning electron microscope image. The target refractive index profile is realized by tuning the size of the air holes against the polymer background. Numerical simulation using commercial software (COMSOL Multiphysics) is performed to compare the reflection for a bumped surface without a cloak (b) and with a cloak (c). The simulation shows the magnetic field component in the z direction at 0.6 THz. Spectra map of bumped surface without a cloak (d) and with a cloak (e) measured using reflection terahertz time-domain spectroscopy. The simulated reflection peak positions from the uncloaked bump sample are marked as white crosses in (d). The measured peak positions match reasonably well with the numerical simulation.

The dashed line shown in (e) illustrates the expected peak position for the cloaked bump sample.

13. Authors thank Jensen Li for fruitful discussions. This work was supported by National Science Foundation under award number CMMI- 0955195, CMMI-0751621, and ECCS-0725764.

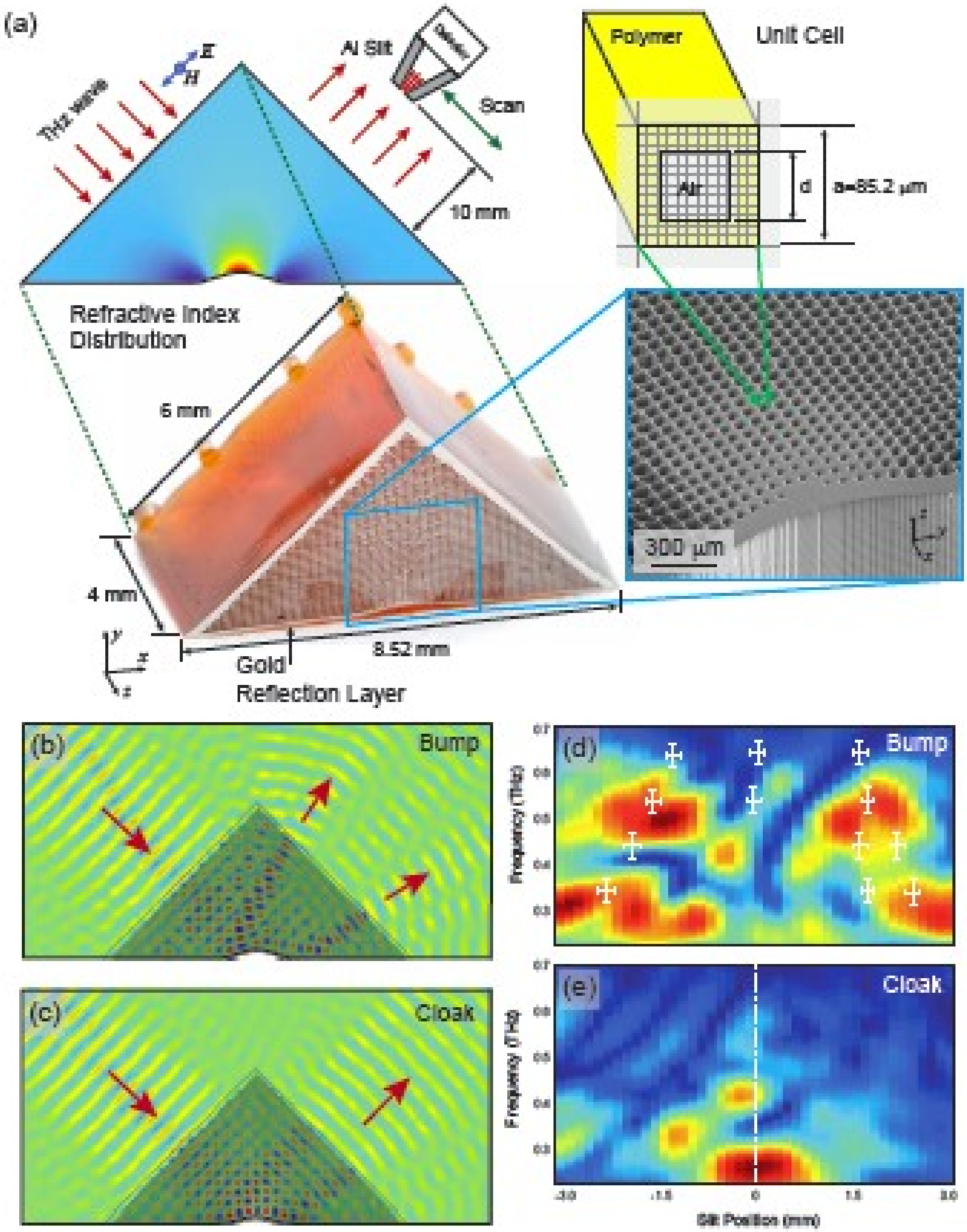

(a)
THz wave
E
H
Al Slit
Scan
10 mm
Polymer
Unit Cell
Air
d
a=85.2 μm
Refractive Index Distribution
6 mm
4 mm
8.52 mm
Gold Reflection Layer
300 μm
(b)
Bump
(c)
Cloak
(d)
Bump
(e)
Cloak
Frequency (THz)
-3.0
-1.5
0
1.5
3.0
Slit Position (mm)